# Change of Optical Properties of Space under Gravitation Field

Vlokh R.

Institute of Physical Optics, 23 Dragomanov Str., 79005 Lviv, Ukraine

## Abstract

It is shown that, in the model of a flat 3D space, the time (i.e., the *Hubble* or the gravitation constant) plays a role of a spatial property. Gravitation field of spherical central mass does not lead to a lowering of symmetry of the space and appearance of anisotropy. In particular, the relations that describe the changes in the refractive indices of the space treated as an optical medium near the massive spherical body under the influence of its gravitation field have been derived.

## Introduction

The idea of anisotropic Universe belongs to *Kantowski and Sachs* [1]. According to this hypothesis, expansion of the Universe is not the same in different directions. Unfortunately, there have not yet been the experimental results which could confirm for certain or reject the *Kantowski*'s *and Sachs*'s model of the Universe. Moreover, as far as we know, the cosmology mainly disposes rather poor experimental facts. Among these facts, one can recall a red shift, a perihelion shift, a radar echo delay and light deflection by massive objects and a background microwave radiation, all of which are known from the basic physics (see, e.g., [2]). However, recent results obtained in different laboratories (see, e.g., [3]) have revealed anisotropy of the background microwave radiation, thus suggesting anisotropy of the local Universe. On the other side, the approach offered by *Kamal, Nandi and Anwarul Islam* [4], *Evans* [5] and *Fernando de Felice* [6] for the description of optical phenomena in gravitation field has been based on optical-mechanical analogy of general relativity and a refractive medium characterized with some effective refractive index. This approach makes an important bridge between the geometric general relativity and physical optics. Furthermore, the authors [5,6] have considered the optical medium, following from the assumptions of its three-dimensionality, flatness, inhomogeneity and isotropy. Besides, the gravitation field is a vectorial one, whose presence could lead to a lowering of symmetry of the space (or the matter). Moreover, inhomogeneous matter should be usually anisotropic.

The present report is devoted to analysis of optical properties of the space induced by the gravitation field.

## Refractive index dependence on the gravitation field

According to [5,6], the coordinate dependence of the refractive index change in the gravitation field of spherical mass may be written as

$$n(r) = \left(1 + \frac{m}{2r}\right)^3 \left(1 - \frac{m}{2r}\right)^{-1}, \quad (1)$$

where $m = \frac{GM}{c_0^2}$, $G$ is the gravitation constant, $r$ the Euclidean radial coordinate, $M$ the spherically symmetric mass and $c_o$ the light velocity for the free space. For example, the effective refraction index calculated on the basis of Eq. (1) for the vicinity of the Sun surface is n=1.000004248. Let us notice that, at first blush, Eq. (1) does not include any material coefficients that could characterize the space and the matter and link the gravitation field strength $g = \frac{GM}{r^2}$ with the refractive index of the space (or the so-called "optical medium", in terms of [5,6]). Upon a closer examination, Eq. (1) could be easily presented as a function $n(g)$ of the gravitation field after introducing $r = \pm\sqrt{\frac{GM}{g}}$ (at this stage we neglect the minus sign, because the radial coordinate can acquire only positive values):

$$n(g) = \left(1 + \frac{\frac{GM}{c_0^2}}{2\sqrt{\frac{GM}{g}}}\right)^3 \left(1 - \frac{\frac{GM}{c_0^2}}{2\sqrt{\frac{GM}{g}}}\right)^{-1} =$$

$$= \left(1 + \frac{(GMg)^{1/2}}{2c_0^2}\right)^3 \left(1 - \frac{(GMg)^{1/2}}{2c_0^2}\right)^{-1}. \quad (2)$$

It is seen already from Eq. (2) that $\frac{(GMg)^{1/2}}{2c_0^2} \ll 1$. On the other hand, the quantity $G$, i.e. the gravitation coefficient, can in general manifest properties of a polar rank-two tensor, since it relates with each other the two polar vectors in the *Newtonian* gravitation equation.

Eq. (2) may be simplified with accounting for that $\frac{(GMg)^{1/2}}{2c_0^2} \ll 1$. Really, this value is of the order of $10^{-5}$, e.g., at the surface of the Sun.

As a result, the refractive index and optical-frequency dielectric impermeability constant $B_{ij} = \left(\frac{1}{n^2}\right)_{ij}$ may be presented as a power series,

$$n = 1 \pm 2\sqrt{\frac{GM}{c_0^4}}(\vec{g}^{1/2})_k + \frac{3}{2}\left(\sqrt{\frac{GM}{c_0^4}}\right)^2 (\vec{g}^{1/2})^2_k \pm$$
$$\pm \frac{1}{2}\left(\sqrt{\frac{GM}{c_0^4}}\right)^3 (\vec{g}^{1/2})^3_k + \frac{1}{16}\left(\sqrt{\frac{GM}{c_0^4}}\right)^4 (\vec{g}^{1/2})^4_k, \quad (3)$$

$$B_{ij} = 1 \mp 4\sqrt{\frac{GM}{c_o^4}}(\vec{g}^{1/2})_k - 7\left(\sqrt{\frac{GM}{c_o^4}}\right)^2 (\vec{g}^{1/2})^2_k \mp$$
$$\mp 7\left(\sqrt{\frac{GM}{c_o^4}}\right)^3 (\vec{g}^{1/2})^3_k - \frac{35}{8}\left(\sqrt{\frac{GM}{c_o^4}}\right)^4 (\vec{g}^{1/2})^4_k. \quad (4)$$

The signs "±" appear in Eqs. (3) and (4) once more in consequence of taking the square root $(\vec{g}^{1/2})_k$. Let us now analyze Eqs. (3), (4). First of all, one can neglect the terms of third and fourth orders, since they are very small. Thus, Eqs.(3) and (4) may be rewritten as

$$n = 1 \pm 2\sqrt{\frac{GM}{c_0^4}}(\vec{g}^{1/2})_k + \frac{3}{2}\left(\sqrt{\frac{GM}{c_0^4}}\right)^2 (\vec{g}^{1/2})^2_k, \quad (5)$$

$$B_{ij} = 1 \mp 4\sqrt{\frac{GM}{c_o^4}}(\vec{g}^{1/2})_k - 7\left(\sqrt{\frac{GM}{c_o^4}}\right)^2 (\vec{g}^{1/2})^2_k. \quad (6)$$

Introducing the notation $\beta_{ijkl} = \left(\sqrt{\frac{G}{c_o^4}}\right)^2 = \frac{G}{c_0^4}$, we rewrite Eqs. (5) and (6) in the form

$$n = 1 \pm 2\sqrt{\beta_{ijkl}M}(\vec{g}^{1/2})_k + \frac{3}{2}\beta_{ijkl}M(\vec{g}^{1/2})^2_k, \quad (7)$$

$$B_{ij} = 1 \mp 4\sqrt{\beta_{ijkl}M}(\vec{g}^{1/2})_k - 7\beta_{ijkl}M(\vec{g}^{1/2})^2_k, \quad (8)$$

where $\beta_{ijkl}$ stands for a fourth-rank polar tensor. It is interesting to recall that the quantity $\frac{G}{c_0^4}$

plays a role of coupling coefficient between the metric tensor (the *Ricci*'s tensor) and the energy-momentum tensor in the general relativity relation. The tensor $\beta_{ijkl}$ represents a material tensor of the space. While assuming the space to be an isotropic matter in its initial state, we may define its symmetry as $\infty/\infty/mmm$. Let us analyze the symmetry properties of $\beta_{ijkl}$, following from the relation $\beta_{ijkl} = \frac{G}{c_0^4}$. Obviously, they depend on the properties of the coefficient $G$, which should be, in general, an asymmetric rank-two polar tensor ($G_{ij} \neq G_{ji}$), with the following limiting properties:

a) a scalar, $G$;

b) a symmetric tensor, $G_{np} = \frac{1}{2}(G_{np} + G_{np}^*)$;

c) an antisymmetric tensor, $G_{np} = \frac{1}{2}(G_{np} - G_{np}^*)$.

In the case (a) we have $\beta_{ijkl} = \frac{G}{(c_{0i} \times c_{0j} \times c_{0k} \times c_{0l})}$, where the denominator represents a dyadic product of four polar vectors and so $\beta_{ijkl}$ is fully symmetric fourth-rank polar tensor with the internal symmetry $[V^4]$.

The symmetric part of the tensor $G_{np}$ can exist in the space with lowered symmetry. Then we have $\beta_{npkl} = \frac{G_{np}}{([c_{0i} \times c_{0j}]_k [c_{0k} \times c_{0l}]_l)}$. In such a case, the form of the tensor and the number of its independent components would depend upon particular symmetry of the space.

The antisymmetric part of $G_{np}$ is an axial vector (pseudo-vector). This vector is nonzero for the space belonging to the symmetry groups, which are subgroups of the axial vector group ($\infty/m$). According to the tensor product rules, one has $\beta_{ijml} = \frac{G_l}{((c_{0i}c_{0j})[c_{0k} \times c_{0l}]_m)}$, where the denominator manifests properties of a third-rank axial tensor. In this case $\beta_{ijml}$ is a fourth-rank polar tensor symmetric in its first three indices (the internal symmetry $[V]^3V$). The antisymmetic part of $G_{pn}$ may be essential if massive rotation singularities exist in the space.

The next question that appears in the analysis of these relations is as follows: what tensorial properties are characteristic of the quantities which include the gravitation field strength $(\vec{g}^{1/2})_k$ and $(\vec{g}^{1/2})_k^2$? Both the quantities $(\vec{g}^{1/2})_k^2 = \sqrt{(g_k \times g_k)} = |g|$ and $(\vec{g}^{1/2})_k = \sqrt[4]{(\vec{g}^{1/2})_k^2} = \pm\sqrt{|g|}$ are scalars. One can therefore come to the conclusion that the gravitation field of spherical mass as the action itself cannot lead to lowering of symmetry of the space. In other words, initially isotropic space with the symmetry $\infty/\infty/mmm$ would remain isotropic under the action of the gravitation field of spherical mass. Moreover, the sign "±" near the odd-power terms correspond to possible opposite signs of the gravitation field. As seen from Eq. (7), in case of a negative gravitation field, the odd-power terms turn out to be negative and might therefore have led to decrease in the refractive index down to the values less than unity, resulting increase of the light velocity up to the values higher than $c_o$.

For the case of spherical massive body, $|g|$ is a scalar, and then the tensor $\beta_{ijml}$ may be convolved into a second-rank tensor $\beta_{ij}$, with the properties:

a) of a scalar;

b) of a symmetric tensor;

c) of an antisymmetric tensor (or, quite equivalent, an axial vector).

Let us now estimate the change of the refractive index (or the light velocity), e.g., for the light beam propagating from the Sun towards the Earth under the gravitation field of Sun. The relevant coefficient is equal to $\beta_{11} = 8.27 \times 10^{-45} \text{s}^2/\text{m} \times \text{kg}$. One can neglect in Eqs. (7) and (8) a very small quadratic term. As a result, these equations yield in

$$n = 1 + 2\sqrt{\beta_{ij} M} (g^{1/2}),  \quad (9)$$

$$B_{ij} = 1 - 4\sqrt{\beta_{ij} M} (g^{1/2}). \quad (10)$$

Taking the solar mass and radius values ($M_s = 1.991 \times 10^{30}$ kg and $R_s = 6.96 \times 10^8$ m), we have $g = 273.98$ m/s$^2$ and so the change in the refractive index induced by the gravitation field of the Sun on its surface becomes $\Delta n = 0.42 \times 10^{-5}$. This very small value is the same as that calculated on the basis of Eq. (1). However, the light path might be large enough at the cosmic scales. The resulting optical retardation can therefore be quite accessible for practical experimental measurements. Besides, the Sun may appear to be not the best example. There exist many other massive objects in the Universe that produce much stronger gravitation fields, such as neutron stars or white drafts.

In the presence of large massive objects, the light beam changes gradually its propagation direction. It is interesting to notice that this well-known effect, which follows from the *Einstein* geometrical general relativity, may be equally well described in frame of the flat space model, with taking the refraction index to be inhomogeneously distributed through the beam cross-section [6]. Therefore, while estimating the total optical path difference occurring for the light passing from a massive body (e.g., from the Sun to the Earth), one can represent it with the relation

$$\Delta = \int_{R_s}^{l} \Delta n(l) dl = 2 \int_{R_s}^{l} \sqrt{\beta_{11} M_s} \sqrt{\frac{M_s G}{(R_s + l)^2}} dl = \\ 2 M_s (\beta_{11} G)^{1/2} \ln \frac{(l + R_s)}{2 R_s}, \quad (11)$$

where $R_s$ denotes the Sun radius and $l = 150 \times 10^9$ m the distance between the Sun and the Earth (the semi-major axis). Taking into account $e = 0.016707$ for $T = J 2004.5$ (see e.g.[7]) $l_{max} = 152.09 \times 10^9$ m (aphelion distance) and $l_{min} = 147.09 \times 10^9$ m (perihelion distance), we derive $\Delta_{max} = 13.897$ km and $\Delta_{min} = 13.798$ km for the mentioned example. It is evident that the optical path difference induced by the gravitation field of the Sun is quite large even for the case of light propagation from the Sun to the Earth. Thus, the delay of light coming from the Sun to the Earth is

$$\Delta t = \frac{\Delta}{c_0}.$$

With $c_0 = 2.99792458 \times 10^8$ m/s it is equal to $\Delta t_{max} = 4.636 \cdot 10^{-5}$ s and $\Delta t_{min} = 4.603 \cdot 10^{-5}$ s as well as $\delta(\Delta t) = \Delta t_{max} - \Delta t_{min} = 3.3 \times 10^{-7}$ s. The effect of the gravitation-induced change of the light speed should become experimentally accessible for the electromagnetic waves passing from satellites to surfaces of planets and back, while employing the measurements of time delay of the returned signal.

Returning to the question of possible optical anisotropy of the space induced with the gravitation field, one can suppose that the gravitation field of non-spherical mass would lead to lowering of the space symmetry, appearance of its anisotropy and so optical birefringence. On the other hand, we may assume that the anisotropy could also appear due to either interaction of the gravitation field with the other fields (e.g., the magnetic one, $\Delta B_{ij} \sim g^{1/2} H_k$) or existence of considerable gradients of the gravitation field (e.g., $\Delta B_{ij} \sim \partial^2 (g)^{1/2} / \partial x_k \partial x_l$).

It is worth noticing that the quantities $\beta$ and $G$ are material coefficients of the flat space (or the corresponding optical medium) and should therefore obey *von Neumann* principle. Hence, lowering of initially spherical symmetry of the space by the gravitation field or the other fields can lead to appearance of tensorial properties of the $G$ coefficient. Moreover, if $\beta$ and $G$ are material coefficients, then it follows, e.g., from the relation for the *Hubble* constant $H^2 = \frac{8}{3} \rho_c \pi G$ (with $\rho_c$ being the critical density of the Universe), that the above constant (because of the relation $H = \frac{1}{t}$, this is true of the time, too) plays a role of property of the flat space in the model of optical medium. The existing fields lead to lowering of the space symmetry, and these lowered groups allow subsequently lowering of symmetry of the properties of the space, i.e., the symmetry of the time (or that of the *Hubble* and the gravitation constants). Therefore, due to the *Neumann* principle, the symmetry group of the flat space should depend on the field configuration and, following the *Curie* symmetry principle, it should be a subgroup of symmetry group of the time.

**Conclusions**

Hence, we have shown in the present paper that the time (as well as the *Hubble* or the gravitation constants) plays a role of spatial property within the model of a flat 3D space. It is assumed that the fields existing in the space could lead to lowering of symmetry of the space and appearance of its anisotropy. The relations that describe changes, due to influence of the gravitation field of spherically symmetric mass, in the refractive indices of the space (taken as an optical medium) have been derived for the vicinity of a massive spherical body. We have also obtained the relation for description of changes in optical impermeability of the space near the spherical mass. It looks as follows:

$$B_{ij} = 1 - 4\sqrt{\beta_{ij} M}(g^{1/2}),$$

where $\beta_{ij} = \dfrac{G}{c_0^4}$ is a second-rank polar tensor representing a material coefficient of the space.

The optical path difference is estimated to be $\Delta_{max} = 13.897\ km$ for the light beam passing from the Sun surface to the Earth (in aphelion) and the corresponding delay of light is equal to $\Delta t_{max} = 4.636 \cdot 10^{-5}$ s. It is shown that initially isotropic space with the symmetry $\infty/\infty/mmm$ should remain optically isotropic under the action of the gravitation field of spherical massive body, though the anisotropy may appear in the gravitation field of the other configurations. This anisotropy allows lowering of the time symmetry from scalar to a second-rank tensor.

The subsequent results on the subject will be reported in a forthcoming paper.